\documentclass[twocolumn,preprintnumbers,amsmath,amssymb,superscriptaddress]{revtex4}
\usepackage{graphicx}
\usepackage{dcolumn}
\usepackage{bm}
\usepackage{soul}
\usepackage{color}
\usepackage{epstopdf}
\usepackage[version=3]{mhchem}
\usepackage{lipsum}
\usepackage[outercaption]{sidecap}
\usepackage{floatrow}
\usepackage{hyperref}
\begin{document}

\title{Cooperative nanoparticle self-assembly and photothermal heating in a flexible plasmonic metamaterial}
\author{Anh D. Phan}
\email{anh.phanduc@phenikaa-uni.edu.vn}
\affiliation{Phenikaa Institute for Advanced Study, Artificial Intelligence Laboratory, Faculty of Computer Science, Materials Science and Engineering, Phenikaa University, Hanoi 12116, Vietnam}
\affiliation{Department of Nanotechnology for Sustainable Energy, School of Science and Technology, Kwansei Gakuin University, Sanda, Hyogo 669-1337, Japan}
\author{Vu D. Lam}
\affiliation{Graduate University of Science and Technology, Vietnam Academy of Science and Technology, 18 Hoang Quoc Viet, Hanoi, Vietnam}
\author{Katsunori Wakabayashi}
\affiliation{Department of Nanotechnology for Sustainable Energy, School of Science and Technology, Kwansei Gakuin University, Sanda, Hyogo 669-1337, Japan}
\date{\today}

\begin{abstract}
We theoretically investigate equilibrium behaviors and photothermal effects of a flexible plasmonic metamaterial composed of aramid nanofibers and gold nanoparticles. The fiber matrix is considered as an external field to reconfigure a nanoparticle assembly. We find that the heating process tunes particle-particle and fiber-particle interactions, which alter adsorption of nanoparticles on fiber surfaces or clustering in pore spaces. Thus, it is possible to control the nanoparticle self-assembly by laser illumination. Gold nanoparticles strongly absorb radiations and efficiently dissipate absorbed energy into heat. By solving the heat transfer equation associated with an effective medium approximation, we calculate the spatial temperature rise. Remarkably, our theoretical results quantitatively agree with prior experiments. This indicates that we can ignore plasmonic coupling effects induced by particle clustering. Effects of the laser spot size and intensity on the photothermal heating are also discussed.
\end{abstract}

\maketitle
\section{Introduction}
Plasmonic metamaterials have attracted a significant attention for a wide range of applications, such as energy harvesting, highly sensitive sensors, high-resolution imaging, and photodetectors \cite{16,14,15,31}. When metal nanostructures are fabricated on flexible and elastomeric substrates or embedded in polymers or fibers, the metamaterials become stretchable, flexible and wearable to surpass limitations of their rigid counterparts, particularly for the biomedical applications \cite{16,15,31,1}. One of the most common polymers used for these metamtaterials is polydimethylsiloxane (PDMS) \cite{16,15,31}. The low surface energy of PDMS generates a hydrophobic surface.

Recently, an ultra-flexible plasmonic metamaterial film is fabricated by mixing gold nanoparticles (AuNPs) and aramid nanofibers \cite{1} to generate broadband optical absorbers with durable mechanical strength. In this metamaterial, the aramid nanofibers have strong mechanical strength with a high flexibility. While localized surface plasmon resonances of gold nanoparticles enhance electromagnetic fields around the nanostructures, absorb light energy and effectively dissipate it into heat when illuminated. Thus, these flexible metamaterials become multifunctional materials for a wide range of applications. However, the photothermal heating increases temperature of the metamaterials, and possibly affects the clustering of nanoparticles and structurally rearranges them. Here, several questions arise: Does the nanoparticle assembly have a significant effect on the photothermal heating? Does the plasmonic coupling matter? Can existing theoretical approaches describe photothermal responses without introducing new physics? The mechanisms have not been comprehensively reported so for. Theoretical insights into microstructures of fiber-nanoparticle composites would  pave the way for applications in various areas, such as drug delivery \cite{10}, neuromorphic transistor \cite{11}, direct ink writing \cite{17}, and hyper-stretchable nanocomposite \cite{18}.

The light-induced temperature increase has been theoretically investigated in various systems. Neumann and her coworkers have simplified the heat diffusion equation to describe qualitatively the heating process and explain how air bubbles around photothermal agents are formed in water \cite{38}. Sousa-Castillo \emph{et. al} have determined thermal gradients of a gold nanocapsule deposited on a glass substrate under simulated solar irradiation \cite{39} by finite element analysis. Recently, we have proposed new models \cite{21, 40} to calculate time-dependent temperature distributions of aqueous solutions of plasmonic nanostructure solutions under solar irradiation. We treat this system as a homogeneous single layer system and capture the collective heating of the nanoparticles for a wide range of particle densities. In the year of 2020, these approaches have been developed to study the thermal gradients of multilayered systems such as graphene metamaterials \cite{30} and floating/hanging PANi fabrics \cite{41}. Numerical results in Ref. \cite{21, 40, 30,41} are in quantitative accordance with experiments. However, photothermal effects in nanoparticle-fiber composites have not been studied yet. The presence of fiber network introduces confinement effects on nanoparticles and possibly shortens the separation distance among them. Thus, strong plasmonic coupling between plasmonic nanoparticles can occur. In addition, while the distribution of nanoparticles in dilute solutions remains unchanged during the heating process, the thermal-induced reorganization of particles in nanoparticle-fiber composites is complicated.

In this work, we use the statistical mechanics theory to study the equilibrium structure and collective light heating of gold nanoparticles dispersed in a disorder nanofiber network using the polymer reference interaction site model (PRISM) integral equation theory. This composite is fabricated as an ultra-flexible plasmonic metamaterial in Ref. \cite{1}. We determine contribution of particle-particle and fiber-particle interactions, and their length scales to the equilibrium isotropic structure. Our approach is useful for explaining other similar experimental systems such as nanoparticle-protein superlattice wires \cite{9}. These interactions can be tuned by laser heating under laser illumination. By using effective medium approximation, we can calculate spatial temperature increase. Our calculations are quantitatively close to experimental results in Ref. \cite{1}. We discuss effects of the nanoparticle density, laser spot, and laser intensity on on the heating process.

\section{Theory and Model}
\subsection{Cooperative nanoparticle self-assembly}
We present a theoretical model for the binary mixture of gold nanoparticles ($p$) dispersed in a randomly distributed nanofiber ($f$) network in Ref. \cite{1} at the interaction site level. Each fiber is modeled as a freely jointed chain (FJC \cite{5}) of $N$ segments or interaction sites with a site diameter (the width of fiber) $d$ and a rigid bond length $l=4d/3$. The persistence length of $4/3$ is a common value used for flexible polymers \cite{6}. Since hydrogen bonds in the fiber network are significantly weakened during fabrication process, we assume that a fiber site interacts with other fiber sites via a hard-core potential model. A gold nanoparticle has an interaction site with a diameter $D$. The interaction between nanoparticles are described by a short-range interaction with a hard-core repulsion, i.e. $u_{pp}(r)=-\varepsilon_{pp}e^{-(r-D)/\delta}$. The interaction between nanoparticles and fibers are also described in the same manner, but a different form, i.e. $u_{fp}(r)=-\varepsilon_{fp}e^{-(r-(D+d)/2)/\delta}$. Here, $\varepsilon_{pp}$ ($\varepsilon_{fp}$) is an interaction strength between nanoparticles (between nanoparticles and fibers). $\delta$ is the parameter for spatial extent of the interaction. One can tune $\varepsilon_{pp}$ and $\varepsilon_{fp}$ via optical techniques \cite{2}, while the spatial range depends on chemical details of systems. 

According to the fabrication in Ref. \cite{1}, the diameter of AuNPs is $D = 0.75d$ = 58 nm, the pore (mesh) size is $\xi \approx 3.23D$, and $N = 20$, and the particle packing fraction $\Phi_p =10^{-3}$ corresponds to the 5.7 wt $\%$ Au content. The pore size is smaller than the length of fiber to construct network. The pore size of the network is theoretically estimated by \cite{3}
\begin{eqnarray}
\xi \approx \left(\frac{3\pi d^2}{4\Phi}\right)^{1/2},
\label{eq:0}
\end{eqnarray}
where $\Phi$ is a packing fraction of the fiber matrix. To obtain $\xi \approx 3.23D$, we have $\Phi = 0.4$.

We compute site-site pair correlation functions for objects using the PRISM theory \cite{19}, or solving numerically Ornstein-Zernike (OZ) equations \cite{4,19}.  In Fourier space, the matrix site-site generalized Ornstein-Zernike equations is given by 
\begin{eqnarray}
h_{ij}(k)=\omega_i(k)\left[C_{ij}(k)\omega_j(k) + \sum_l C_{il}(k)\varrho_l h_{lj}(k) \right],
\label{eq:a1}
\end{eqnarray}
where $k$ is wavevector, $h_{ij}(r)= g_{ij}(r) - 1$, $g_{ij}(r)$ is the site-site radial distribution function between species $i$ and $j$, and $C_{ij}(r)$ is the direct correlation function, $\varrho_i$ is the site number density of species $i$, and $\omega_j(k)$ is the single molecule structure factor of species $j$. For convenience, subscript $"p"$ indicates particles and $"f"$ indicates fibers. The radial distribution functions quantify the probability of a tagged particle to find another particle at distance $r$.

For gold particles, $\omega_p(k) = 1$. Meanwhile, the molecule structure factor of FJC polymer is \cite{6}
\begin{eqnarray}
\omega_f(k)=\frac{1}{(1-f)^2}\left[1-f^2-\frac{2f}{N}+\frac{2f^{N+1}}{N} \right],
\label{eq:a2}
\end{eqnarray}
where $f =\sin(4kd/3)/(4kd/3)$.

To solve Eq.(\ref{eq:a1}), we apply the Percus-Yevick (PY) closure \cite{4,19} to the standard fiber-fiber and fiber-nanoparticle real space correlations. The site-site PY approximation for correlations inside species is zero. For outside species, this is
\begin{eqnarray}
C_{ij}(r) = \left(1-e^{u_{ij}(r)/k_BT}\right)g_{ij}(r),
\label{eq:a3}
\end{eqnarray}
where $k_B$ is the Boltzmann constant and $T$ is the ambient temperature. The closure for nanoparticle-nanoparticle correlation is the hypernetted-chain (HNC) closure \cite{4,19}, which is
\begin{eqnarray}
C_{pp}(r) = h_{pp}(r)-\ln g_{pp}(r)-\frac{u_{pp}(r)}{k_BT}.
\label{eq:a4}
\end{eqnarray}
Now, one can employ the Picard or a Newton-Raphson algorithm \cite{4} to numerically solve Eqs. (\ref{eq:a1})-(\ref{eq:a4}) to obtain the radial distribution functions and structure factors.

\subsection{Photo-thermal heating}
Although pristine aramid nanofibers has a relatively low optical absorption ($\sim 20$ $\%$), this absorption is significantly enhanced by embedding gold nanoparticles. It was experimentally proved that the AuNPs-nanofiber composite absorbs nearly 100 $\%$ of incident light in the wavelength range from 400 nm to 2000 nm when $\Phi_p \geq 2.3\times 10^{-3}$ \cite{1}. The absorbed energy thermally dissipates into the system, raises temperature, and reduces particle-particle and fiber-particle interactions. 

Under broadband laser irradiation, the spatial temperature gradient is analytically obtained by solving a heat energy balance equation of a slab. The temperature rise at coordinate ($x$, $y$, $z$) \cite{20,21} is
\begin{widetext}
\begin{eqnarray}
\Delta T(x,y,z,t)=\frac{I(1-\mathcal{R})\alpha}{2\rho_m c_m}\int_0^t \exp\left(-\frac{\beta^2(x^2+y^2)}{1+4\beta^2\kappa t'} \right)\frac{e^{\alpha^2\kappa t'}}{1+4\beta^2\kappa t'}\left[e^{-\alpha z}\ce{erfc}\left(\frac{2\alpha\kappa t'-z}{2\sqrt{\kappa t'}} \right)+ e^{\alpha z}\ce{erfc}\left(\frac{2\alpha\kappa t'+z}{2\sqrt{\kappa t'}} \right)\right]dt',
\label{eq:1}
\end{eqnarray}
\end{widetext}
where $xy$ plane is in-plane and $z$ is the depth direction. $I$ is the laser intensity, $\beta$ is the inverse of the laser spot radius, and $\mathcal{R}$ is the reflectivity. Since the ultra-flexible metamaterial has strong anti-reflection, $\mathcal{R} \approx 0$. $\kappa = K_m/(\rho_m c_m)$ is the thermal diffusivity, $K_m$ is the thermal conductivity, $\rho_m$ is the mass density, $c_m$ is the specific heat, and $\alpha$ is the absorption coefficient of the metamaterial. To compare theoretical calculations with photo-thermal characterizations in Ref. \cite{1}, we perform with an incident beam having diameter of 4 mm, $\Phi_p=2.3\times10^{-3}$, and $I$ = 30.4 \ce{mW/mm^2}. 

According to effective medium approximations \cite{22,23}, an effective dielectric function of the metamaterials is $\epsilon_{eff}=\epsilon_f + \Phi_p\epsilon_p$, here $\epsilon_f$ and $\epsilon_f$ being the dielectric function of the aramid fiber \cite{1} and gold nanoparticles \cite{29}, respectively. We assume that the presence of nanoparticles does not change the packing fraction and macroscopic structure of the aramid fiber. Gold nanoparticles simply fill in the pore space. From this, one can determine the absorption coefficient by
\begin{eqnarray}
\alpha = \frac{4\pi}{\lambda}\ce{Im}\left(\sqrt{\epsilon_{eff}} \right) = \frac{4\pi}{\lambda}\ce{Im}\left(\sqrt{\epsilon_f + \Phi_p\epsilon_p} \right),
\label{eq:5}
\end{eqnarray}
where $\lambda$ is the wavelength of incident light.

The thermal conductivity, mass density, and specific heat of the metamataterial can be approximately calculated using effective medium approximations \cite{22,23,24}
\begin{eqnarray}
K_m &=& K_{f}+\Phi_pK_{p}, \nonumber\\
\rho_m &=& \rho_{f} + \Phi_p\rho_{p}, \nonumber\\
c_m &=& c_{f} + \Phi_pc_{p}, 
\label{eq:2}
\end{eqnarray}
where the subscript $f$ and $p$ denote to the corresponding properties of the pure aramid fiber network and gold nanoparticles, respectively. Here, $K_{f} = 0.048 $ \ce{Wm^{-1}K^{-1}} \cite{1}, $\rho_{f} = 1440$ \ce{kg/m^3} \cite{25}, $\rho_{p} = 19320$ \ce{kg/m^3}, $c_{f} = 1420$ \ce{J/kg^0C} \cite{25}, and $c_{p} = 129$ \ce{J/kg^0C}. The lattice thermal conductivity depends on the particle size since the phonon scattering is driven by finite-size and boundary effects \cite{26,27}. A previous study \cite{24} reveals that the thermal conductivity of a plasmonic nanoparticle of radius 58 nm is reduced by that of its bulk counterpart. Thus, we use $K_{p} = 115$ \ce{Wm^{-1}K^{-1}} in this work.

\section{Results and discussions}
\subsection{Hard sphere nanoparticles in the interacting fiber network}
\begin{figure}[htp]
\includegraphics[width=9cm]{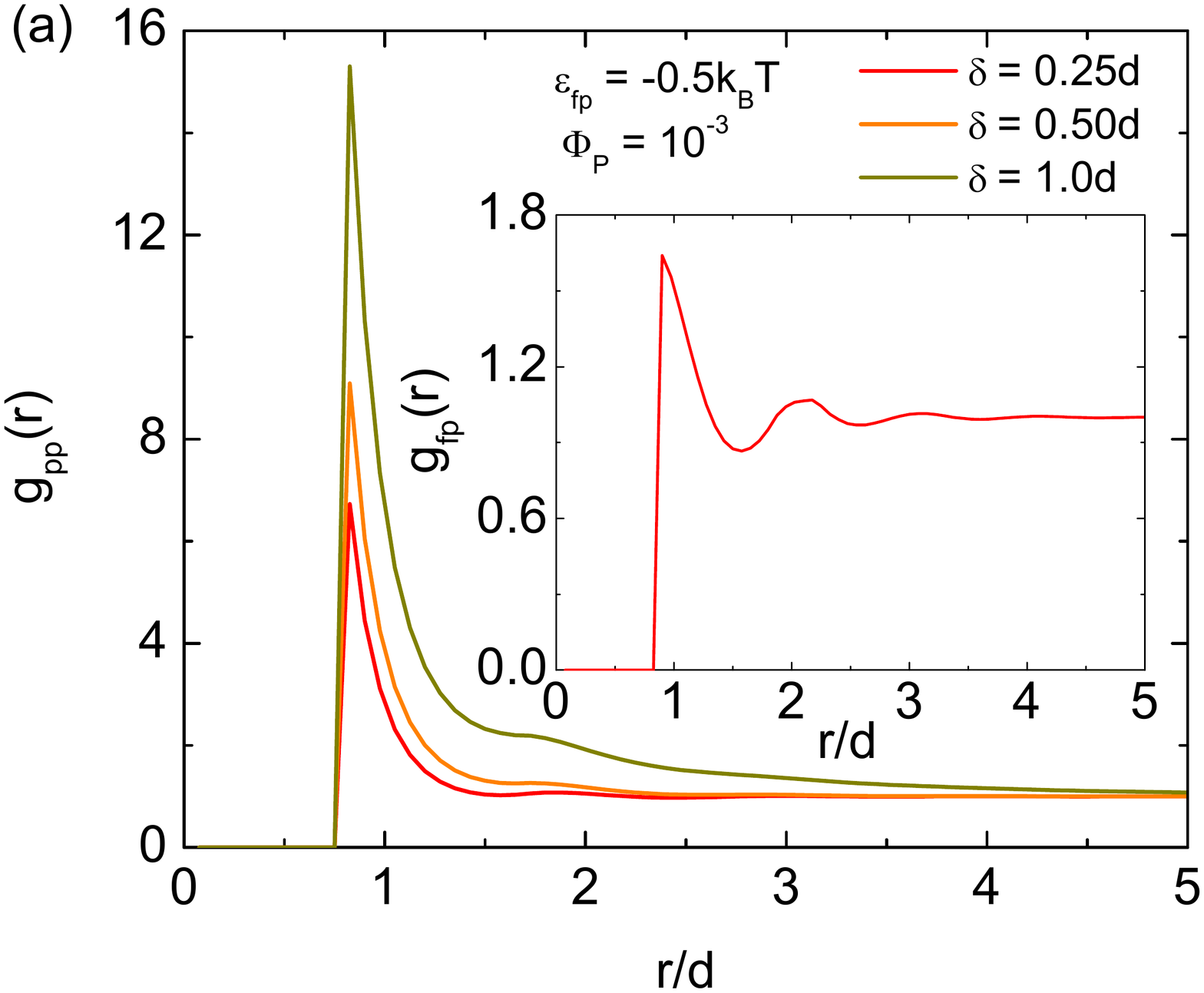}
\includegraphics[width=7cm]{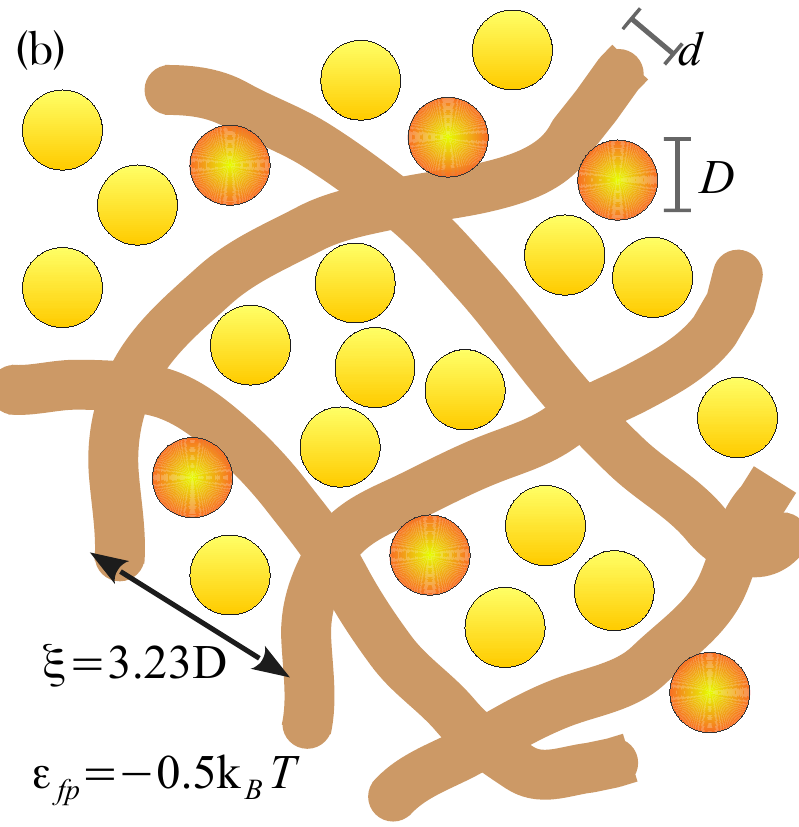}
\caption{\label{fig:1}(Color online) (a) Radial distribution functions in a nanoparticle-fiber mixture for nanoparticle-nanoparticle (mainframe) and nanoparticle-fiber (inset)  pair correlations at $\varepsilon_{fp}=-0.5k_BT$. Numerical results in the inset are calculated at $\delta=0.25d$. (b) The corresponding schematic of the disorder nanofiber network in an equilibrium configuration. Orange and yellow spheres correspond to particles adsorbing on the network or moving in the pore, respectively. Key parameters for sizes of fiber, nanoparticle, and mesh are indicated.}
\end{figure}

Figure \ref{fig:1}a shows particle-particle and fiber-particle radial distribution functions for nanoparticles dispersed in a fiber network with the moderate repulsion of $\varepsilon_{fp}=-0.5k_BT$ and $\varepsilon_{pp}=0$ (hard-sphere repulsion). A single peak of $g_{pp}(r)$ and $g_{fp}(r)$ indicates localization of particles in the small pore space. Meanwhile, since the maximum of $g_{fp}(r)$ is smaller than 1.8, several particles form a very dilute monolayer onfibers. The deduced static microstructure is schematically depicted in Fig. \ref{fig:1}b. 

The particle localization is strongly driven by a spatial range of interfacial forces. As seen in Fig. \ref{fig:1}a, $g_{pp}(r)$ increases with a growth of $\delta$. One observes a tighter localization with the increase of the range of repulsion. This interacting length scale is tunable by varying  a salt concentration. According to the Debye-H\"uckel theory \cite{28}, one has  
\begin{eqnarray}
\frac{1}{\delta} = \sqrt{\frac{2z^2e^2n}{\varepsilon_r\varepsilon_0k_BT}},
\label{eq:3}
\end{eqnarray}
where $\varepsilon_0$ is the vacuum permittivity, $\varepsilon_r$ is the medium permittivity, and $z$ and $n$ are the ionic valence and the ion concentration in solution, respectively.

Equation (\ref{eq:3}) indicates that the increase of salt screens particle-fiber repulsive interactions. In the absence of fiber network or in the large pore-size system, the state of hard-sphere fluids becomes localized as $\Phi_p \geq 0.432$ \cite{4}. However, our results exhibit an emergence of the localized state at $\Phi_p=10^{-3}$ in the presence of a small pore size network. The behavior reveals effects of geometrical constraints on forming the glass-like structure. The localization of particles are stronger at low salt concentrations and weaker (particles start delocalized) at large salt concentrations. These calculations are completely consistent with experiments in Ref. \cite{9}. Although the prior work \cite{9} investigates the assembly of gold nanoparticles and anionic tobacco mosaic virus nanorods, parameters for interactions, density, and sizes for PRISM calculations are relatively similar.

\subsection{Interacting nanoparticles in a non-adsorbing fiber network}
\begin{figure}[htp]
\includegraphics[width=9cm]{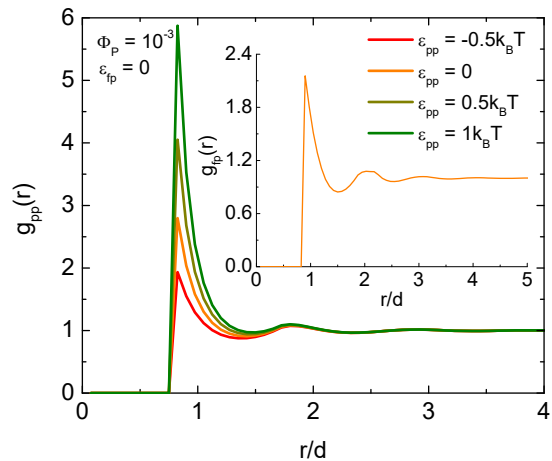}
\caption{\label{fig:5}(Color online) Particle-particle radial distribution functions in the non-interacting fiber network calculated at several values of $\varepsilon_{pp}$ and $\delta =0.25d$. Inset: a nanoparticle-fiber radial distribution function at $\varepsilon_{pp}=0$.}
\end{figure}

The particle-particle interaction also has a significant influence on the structural assembly. We consider equilibrium structure of gold nanoparticles with different values of $\varepsilon_{pp}$ in a neutral fiber matrix ($\varepsilon_{fp}=0$). Numerical results of pair correlation functions are shown in Fig. \ref{fig:5}. Since the function $g_{pp}(r)$ shows weak response to a change of the interacting length scale, we only calculate for $\delta=0.25d$. In purely repulsive particle systems ($\varepsilon_{pp}=-0.5k_BT$ and 0 or hard-sphere repulsion), particles are localized to form particle cages due to geometrical constraints. The caging force originates from the repulsion, which leads to the glass-like structure. When the particle-particle interaction is attractive with a small amount of $\varepsilon_{pp}=0.5 k_BT$, particles stick to each other both in pore and on the fiber network. This is a gel-like state formed by attractive physical bond formation. As the attraction strength is increased more ($\varepsilon_{pp}=1 k_BT$), the number of interparticle contacts or $g_{pp}(D)$ increase. The gelation is stronger.

Interestingly, the spatial fiber-particle correlation is very insensitive to $\varepsilon_{pp}$. The radial distribution function $g_{fp}(r)$ is shown in the inset of Fig. \ref{fig:5}. It implies that interfacial adsorption is unaffected by the interparticle interaction but gold nanoparticles are reorganized to favor either clustering in the pore or sequential multilayer adsorption.

\subsection{Laser induced temperature increase of ultra-flexible metamaterials}
Our results in previous subsections indicate that equilibrium structures are basically insensitive to $\delta$ except for soft-repulsive fiber networks but are sensitive to either the interaction strength $\varepsilon_{pp}/k_BT$ or $\varepsilon_{fp}/k_BT$. Particles can contact to each other in the pore space or on the nanofiber network. The strong plasmonic coupling can occur as reducing interparticle separation distances. Very tight confinement of electromagnetic fields at a gap between a pair of coupled metal nanoparticles leads to emergence of hot spots at the nanojunction interfaces \cite{32,33}. In addition, optical spectra of the coupled structures are significantly shifted compared to those of the isolated counterparts. 

Clearly, arrangement of particles can be controlled by heating. Under laser illumination, gold nanoparticles absorb electromagnetic fields or optical energy of incident light via plasmonic properties. The absorbed optical energy is effectively converted into heat and increases the temperature. The nanoparticle assembly can be continuously changed with turning on and off laser light as a reversible process.

\begin{figure*}[htp]
\includegraphics[width=18cm]{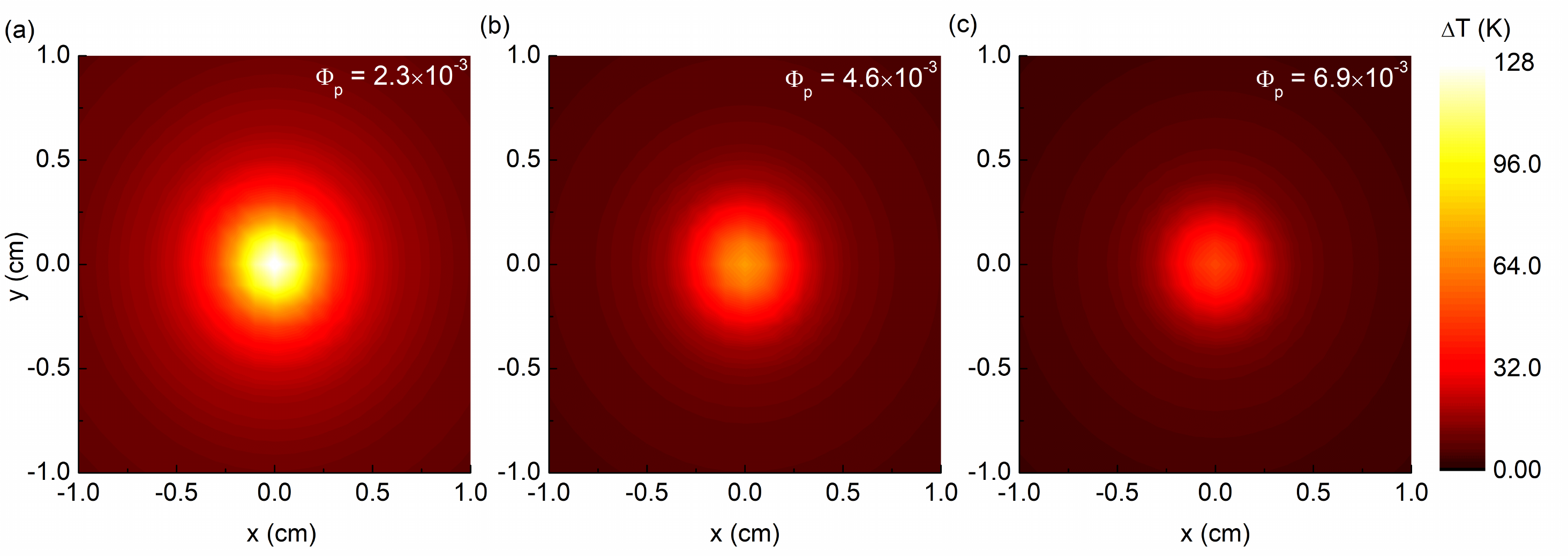}
\caption{\label{fig:6}(Color online) Spatial contour plot of the steady-state temperature increase in Kelvin units on the surface of the ultra-flexible plasmonic metamaterial exposed by a laser spot of 4 mm with the intensity $I$ = 30.4 \ce{mW/mm^2}. The packing fraction of nanoparticle is (a) $\Phi_p=2.3\times10^{-3}$, (b) $\Phi_p=4.6\times10^{-3}$, and (c) $\Phi_p=6.9\times10^{-3}$.}
\end{figure*}

Figure \ref{fig:6}a shows the temperature profile of the steady state under laser illumination with the intensity $I$ = 30.4 \ce{mW/mm^2}. The packing fraction is $\Phi_p=2.3\times10^{-3}$ to mimic photothermal experiments in Ref. \cite{1}. The temperature rise at the surface center is $\Delta T(x=0,y=0) = 128$ $^0C$. In addition, the diameter of the hot area defined by $\Delta T(x,y) \geq 96$ $^0C$ is about 5 mm. These results and our spatial contour plot are very close to the prior study \cite{1}. This agreement between the calculations and experiments validates an effective medium approximation used in Eqs. (\ref{eq:5}) and (\ref{eq:2}). It means that the distribution of gold nanoparticles can be considered as a random dispersion. Although effects of interparticle plasmon coupling are expected to appear when gold nanoparticles are thermally reorganized, the collective heating is insensitive to this coupling enhancement. A possible reason for this behavior is the plasmonic coupling is very weak at wavelength of laser excitation. Another reason may be $\varepsilon_{pp}$ and $\varepsilon_{fp}$ $\ll k_BT$. Consequently, the nanoparticle assembly remains unchanged during a heating process. This explains why multiple-switching light source does not change optical and thermal response of the metamaterial in Ref. \cite{1}.  

According to Eq. (\ref{eq:1}), the temperature rise $\Delta T$ is linearly grows with the laser intensity. This clearly explains data in prior works \cite{1,12,13,30}. Our linear relation is $\Delta T(x=0,y=0)\approx 4.28 I$ , here $I$ is in the unit of \ce{mW/mm^2}. This calculation also quantitatively agrees with experimental results in Ref. \cite{1}.

Figure \ref{fig:6}b and \ref{fig:6}c show the density plots of $\Delta T$ for higher concentrations of nanoparticles ($\Phi_p=4.6\times10^{-3}$ and $6.9\times10^{-3}$) illuminated by laser light. An increase of the volume fraction of gold particles significantly localizes the absorbed optical energy at the interface. However, since the thermal conductivity of the metamaterial grows with increasing $\Phi_p$, the light-to-heat conversion is efficiently conducted. As a result, the laser-induced temperature rise is decreased with densifying nanoparticles. This finding indicates that at the same packing fraction, replacing gold nanoparticles with TiN nanoparticles leads to a higher temperature rise since optical properties of TiN nanoparticles are similar to those of Au counterparts \cite{21} but $K_{TiN}^{bulk}\approx 29$ $Wm^{-1}K^{-1}$ $\ll K_p$ \cite{34}.

\begin{figure}[htp]
\includegraphics[width=9cm]{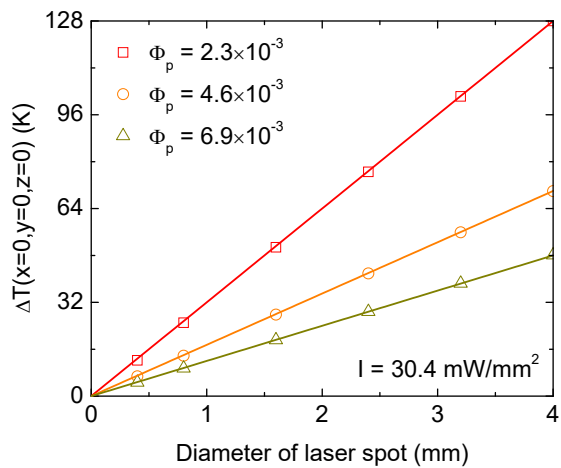}
\caption{\label{fig:7}(Color online) The steady-state temperature rise in Kelvin units of the plasmonic metamaterial at several values of $\Phi_p$ as a function of the laser spot size with $I$ = 30.4 \ce{mW/mm^2}. Open data points and solid lines are our numerical calculations and guide-to-the-eye lines, respectively.}
\end{figure}

Based on Eq. (\ref{eq:1}), we determine the dependence of $\Delta T(x=y=z=0)$ on the light spot size. Figure \ref{fig:7} shows a linear relation between the steady temperature of center point on the surface and the diameter of illuminated laser spot at $\Phi_p=2.3\times10^{-3}$, $4.6\times10^{-3}$, and $6.9\times10^{-3}$. A decrease of the incident laser spot area linearly reduces the temperature increase. Thus, except for the intensity $I$, one can tune $\Delta T$ by varying the diameter of incident light.

There are many other designs for broadband near-perfect absorbers or anti-reflectors. In prior works, the structures are composed of a periodic array of dielectric micro-spheres on a metal film \cite{35}, a periodic array of truncated metal/dielectric cones on a metal substrate \cite{36}, and a three-layer metallo-dielectric-metallo stack \cite{37}. These metamaterials anti-reflect almost all of incident electromagnetic fields and these are similar to our studied metamaterial. However, they are rigid and not flexible. In addition, the thermal conductivity of metal is much larger than that of fiber or even fiber mixed with plasmonic nanoparticles. Thus, under the same laser irradiation, the temperature rise of the rigid plasmonic metamaterials is smaller since heating the system of larger thermal conductivity needs more thermal energy.
\section{Conclusions}
We have investigated equilibrium structures and plasmonic heating of ultra-flexible plasmonic metamaterial including gold nanoparticles confined in aramid nanofiber network. To validate our theoretical approach, we compare with experiments \cite{1}. Experimental information needed for theoretical calculations includes (i) the diameter of nanoparticles, (ii) the diameter and length of each fiber, (iii) either the volume fraction or the mesh/pore size of the fiber matrix, (iv) the volume fraction of nanoparticles that can be calculated using weight-weight percentages, and (v) photothermal heating results. By using the PRISM theory, we have determined the dependence of reconfigurable nanoparticle assemblies on temperature, the strengths of particle-particle and fiber-particle force, and their spatial ranges. Since the pore size is approximately equal to 3 times of particle diameter, the predicted radial distribution function reveals that the separation distance among nanoparticles is significantly shortened and this is strongly temperature-dependent. Gold nanoparticles can adsorb on the fiber network or cluster in the pore space. The reorganization of nanoparticles can lead to plasmonic coupling effects, which are important for the photothermal response of the nanoparticle-fiber composite.

To understand thermal-induced reassembly and plasmonic interactions of gold nanoparticles in this ultra-flexible plasmonic metamaterial, we have theoretically calculated the temperature gradient of the metamaterial under laser illumination. Our approach uses the analytical solution of the heat transfer differential equation associated with an effective medium approximation. This treatment means we ignore enhancement of electromagnetic fields at nanogaps among plasmonic nanostructures. Theoretical temperature rise agrees quantitatively well with prior photothermal characterization \cite{1}. This agreement indicates the plasmonic coupling effect is possibly small to be ignored even when the assembly of gold nanoparticles can be thermally reconfigured. Another possibility is that the fiber-particle and particle-particle interactions are much smaller than $k_BT$ and the structure of nanocomposites is nearly independent of temperature. Our work also suggests it is possible to use a random distribution model to effectively describe the dispersion of nanoparticles in fiber matrix in the same manner as nanoparticles dispersed in aqueous solution. This theoretical approach can be exploited for the rational design of materials with tailored functions. 

\begin{acknowledgments}
This work was supported by JSPS KAKENHI Grant Numbers JP19F18322 and JP18H01154. This research is also funded by Vietnam Academy of Science and Technology under the grant number NVCC42.03/20-20 and KHCBVL.01/20-21.

Conflict of Interest: The authors declare that they have no conflict of interest.
\end{acknowledgments}

\end{document}